\documentclass[12pt,draftcls,onecolumn]{IEEEtran}

\newtheorem{lemma}{Lemma}

\newtheorem{definition}{Definition}
\newtheorem{theorem}{Theorem}
\newtheorem{corollary}{Corollary}

\usepackage[USenglish]{babel}

\usepackage{makeidx}
\usepackage{amsfonts}
\usepackage{epsfig}
\usepackage{amsmath}
\usepackage{amssymb}
\newif\ifcomment\commentfalse

\def\commentOFF{\commentfalse}

\long\outer\def\bc#1\ec{{\ifcomment \sloppy  $[${\bf suggest}]
{{#1}} \textbf{[end]} \fi }}

\long\outer\def\br#1\er{{\ifcomment \sloppy  $[${\bf suggest remove}]
{{#1}} \textbf{[end]} \fi }}

\long\outer\def\bo#1\eo{{\ifcomment \sloppy  $[${\bf instead of}]
{\textit{#1}} \textbf{[end]}  \fi }}

\long\outer\def\BC#1\EC{{\ifcomment \sloppy \par \#  \dotfill
{\textsc{#1}} \dotfill \# \par \fi }}

 \long\outer\def\phmin#1{$PH^{nt}(*,{#1})$}

 \long\outer\def\mpphmin#1{$MPPH^{nt}(*,{#1})$}

\long\outer\def\phminZ{$PH^{nt}$} \long\outer\def\mpphminZ{$MPPH^{nt}$}

\long\outer\def\lbmidmin#1#2{\ensuremath{LB^{nt}_{mid}({#1},{#2})}}
\long\outer\def\lbwith#1#2{\ensuremath{LB_{mid}({#1},{#2})}}

\commentOFF

\ifcomment
\fi

\begin{document}


\title{Shorelines of islands of tractability: Algorithms for parsimony and minimum perfect phylogeny haplotyping problems\thanks{Supported by the Dutch
BSIK/BRICKS
project.}}


\author{Leo van Iersel, Judith Keijsper, Steven Kelk and Leen Stougie}


\maketitle

\begin{abstract}
\noindent The problem \emph{Parsimony Haplotyping} ($PH$) asks for the smallest set of haplotypes which can explain a
given set of genotypes, and the problem \emph{Minimum Perfect Phylogeny Haplotyping} ($MPPH$) asks for the smallest
such set which also allows the haplotypes to be embedded in a \emph{perfect phylogeny}, an evolutionary tree with
biologically-motivated restrictions. For $PH$, we extend recent work by further mapping the interface between ``easy''
and ``hard'' instances, within the framework of $(k,\ell)$-\emph{bounded instances} where the number of 2's per column
and row of the input matrix is restricted. By exploring, in the same way, the tractability frontier of $MPPH$ we
provide the first concrete, positive results for this problem, and the algorithms underpinning these results offer new
insights about how $MPPH$ might be further tackled in the future. In addition, we construct for both $PH$ and $MPPH$
polynomial time approximation algorithms, based on properties of the columns of the input matrix. We conclude with an
overview of intriguing open problems in $PH$ and $MPPH$.
\end{abstract}


\section{Introduction}
\noindent The computational problem of inferring biologically-meaningful haplotype data from the genotype data of a population
continues to generate considerable interest at the interface of biology and computer science/mathematics. A popular
underlying abstraction for this model (in the context of diploid organisms) represents a genotype as a string
over a $\{0,1,2\}$ alphabet, and a haplotype as a string over $\{0,1\}$. The exact goal depends on the
biological model being applied but a common, minimal algorithmic requirement is that, given a set of genotypes, a
set of haplotypes must be produced which resolves the genotypes.
\medskip

To be precise, we are given a \emph{genotype matrix} $G$ with elements in $\{0,1,2\}$, the rows of which
correspond to genotypes, while its columns correspond to sites on the genome, called SNP's. A \emph{haplotype matrix}
has elements from $\{0,1\}$, and rows corresponding to haplotypes. Haplotype matrix $H$ \emph{resolves} genotype
matrix $G$ if for each row $g_i$ of $G$, containing at least one $2$, there are two rows $h_{i_1}$ and $h_{i_2}$ of
$H$, such that $g_i(j) = h_{i_1}(j)$ for all $j$ with $h_{i_1}(j)= h_{i_2}(j)$ and $g_i(j) = 2$ otherwise, in which
case we say that $h_{i_1}$ and $h_{i_2}$ resolve $g_i$, we write $g_i=h_{i_1}+h_{i_2}$, and we call $h_{i_1}$ the {\em
complement} of $h_{i_2}$ with respect to $g_i$, and vice versa. A row $g_i$ without 2's is itself a haplotype and is
uniquely resolved by this haplotype, which thus has to be contained in $H$.

We define the first of the two problems that we study in this paper.

\medskip

\noindent \textbf{Problem:} Parsimony Haplotyping ($PH$)\\
\textbf{Input:} A genotype matrix $G$.\\
\textbf{Output:} A haplotype matrix $H$ with a minimum number of rows that resolves $G$.

\medskip

\noindent \looseness=-2 There is a rich literature in this area, of which recent papers such as \cite{brown} give a
good overview. The problem is APX-hard \cite{lanciaApx}\cite{islands} and, in terms of approximation algorithms with
performance \emph{guarantees}, existing methods remain rather unsatisfactory, as will be shortly explained. This has
led many authors to consider methods based on Integer Linear Programming (ILP)
\cite{brown}\cite{gusfieldparsimony}\cite{halldorson}\cite{lanciaApx}. A different response to the hardness is to
search for ``islands of tractability'' amongst special, restricted cases of the problem, exploring the frontier
between hardness and polynomial-time solvability. In the literature available in this direction
\cite{wabi}\cite{lanciaApx}\cite{lancia}\cite{islands}, this investigation has specified classes of
$(k,\ell)$-\emph{bounded instances}: in a $(k,\ell)$-\emph{bounded instance} the input genotype matrix $G$ has at most
$k$ $2$'s per row and at most $\ell$ $2$'s per column (cf. \cite{islands}). If $k$ or $\ell$ is a ``$*$'' we mean
instances that are bounded only by the number of $2$'s per column or per row, respectively. In this paper we
supplement this ``tractability'' literature with mainly positive results, and in doing so almost complete the bounded
instance complexity landscape.

Next to the $PH$ problem we study the \emph{Minimum Perfect Phylogeny Haplotyping} ($MPPH$)
model \cite{nphardnote}. Again a minimum-size set of resolving haplotypes is required but this time under the
additional, biologically-motivated restriction that the produced haplotypes permit a \emph{perfect phylogeny}, i.e.,
they can be placed at the leaves of an evolutionary tree within which each site mutates at most once. Haplotype
matrices admitting a perfect phylogeny are completely characterised \cite{gusfieldbook}\cite{gusfieldnetwork} by the
absence of the forbidden submatrix\\
\[F = \begin{bmatrix} 1 & 1 \\ 0 & 0 \\ 1 & 0 \\ 0 & 1 \end{bmatrix}.\] \\
\noindent
\textbf{Problem:} Minimum Perfect Phylogeny Haplotyping ($MPPH$)\\
\textbf{Input:} A genotype matrix $G$.\\
\textbf{Output:} A haplotype matrix $H$ with a minimum number of rows that resolves $G$ and admits a perfect
phylogeny.

\medskip

\noindent The feasibility question ($PPH$) - given a genotype matrix $G$, find any haplotype matrix $H$ that resolves
$G$ and admits a perfect phylogeny, or state that no such $H$ exists - is solvable in linear-time
\cite{gusfieldlinear}\cite{anOptimal}. Researchers in this area are now moving on to explore the $PPH$ question on
phylogenetic \emph{networks} \cite{gusnetwork}.

The $MPPH$ problem, however, has so far hardly been studied beyond an NP-hardness result \cite{nphardnote}
and occasional comments within $PH$ and $PPH$ literature \cite{mpphref2}\cite{anOptimal}\cite{mpphref}. In this paper we
thus provide what is one of the first attempts to analyse the parsimony optimisation criteria within a well-defined
and widely applicable biological framework. We seek namely to map the $MPPH$ complexity landscape in the same way as
the $PH$ complexity landscape: using the concept of $(k,\ell)$-boundedness. We write $PH(k,\ell)$ and $MPPH(k,\ell)$ for these
problems restricted to $(k,\ell)$-bounded instances.\\

\noindent
\textbf{Previous work and our contribution}

\medskip

\noindent In \cite{lanciaApx} it was shown that $PH(3,*)$ is APX-hard. In \cite{wabi}\cite{lancia} it was shown that
$PH(2,*)$ is polynomial-time solvable. Recently, in \cite{islands}, it was shown (amongst other results) that
$PH(4,3)$ is APX-hard. In \cite{islands} it was also proven that the restricted subcase of $PH(*,2)$ is polynomial-time solvable
where the \emph{compatibility graph} of the input genotype matrix is a clique. (Informally, the compatibility graph shows
for every pair of genotypes whether those two genotypes can use common haplotypes in their resolution.)

In this paper, we bring the boundaries between hard and easy classes closer by showing that $PH(3,3)$ is APX-hard and
that $PH(*,1)$ is polynomial-time solvable.

As far as $MPPH$ is concerned there have been, prior to this paper, no concrete results beyond the above mentioned
NP-hardness result. We show that $MPPH(3,3)$ is APX-hard and that, like their $PH$ counterparts, $MPPH(2,*)$ and
$MPPH(*,1)$ are polynomial-time solvable (in both cases using a reduction to the $PH$ counterpart). We also show that
the clique result from \cite{islands} holds in the case of $MPPH(*,2)$ as well. As with its $PH$ counterpart the
complexity of $MPPH(*,2)$ remains open.

\medskip
\noindent The fact that both $PH$ and $MPPH$ already become $APX$-hard for $(3,3)$-bounded instances means that,
in terms of deterministic approximation algorithms, the best that we can in general hope for is constant
approximation ratios. Lancia et al \cite{lanciaApx}\cite{lancia} have given two separate approximation algorithms with approximation
ratios of $\sqrt{n}$ and $2^{k-1}$ respectively, where $n$ is the number of genotypes in the input, and $k$ is the maximum
number of 2's appearing in a row of the genotype matrix\footnote{It would
be overly restrictive to write $PH(k,*)$ here
because their algorithm runs in polynomial time even if $k$ is not a constant.}. An $O(\log n)$
approximation algorithm has been given in \cite{log} but this only runs in polynomial time if the set of all possible haplotypes that
can participate in feasible solutions, can be enumerated in polynomial time. The obvious problem with the $2^{k-1}$ and the
$O(\log n)$ approximation algorithms is thus that either the accuracy decays exponentially (as in the former case) or the running
time increases exponentially (as in the latter case) with an increasing number of 2's per row. Here we offer a
simple, alternative approach which achieves (in polynomial time) approximation ratios linear in $\ell$ for
$PH(*,\ell)$ and
$MPPH(*,\ell)$ instances, and
actually also achieves these ratios in polynomial time when $\ell$ is not constant. These ratios are
shown in the Table \ref{tab:ratios}; note how improved
ratios can be obtained if every genotype is guaranteed to have at least one 2.
\begin{table}
\centering
\caption{Approximation ratios achieved in this paper}
\label{tab:ratios}
\begin{tabular}{|c|c|}
\hline
Problem $(\ell \geq 2)$ & Approximation ratio\\
\hline
\hline
$PH(*,\ell)$ & $\frac{3}{2}\ell + \frac{1}{2}$\\
\hline
$PH(*,\ell)$ where every genotype has at least one 2 & $\frac{3}{4}\ell + \frac{7}{4} - \frac{3}{2}\frac{1}{\ell +1}$\\
\hline
$MPPH(*,\ell)$ & $2 \ell$\\
\hline
$MPPH(*,\ell)$ where every genotype has at least one 2 & $\ell + 2 - \frac{2}{\ell+1}$\\
\hline
\end{tabular}
\end{table}

We have thus decoupled the approximation ratio from the maximum number of 2's per row, and instead made the ratio
conditional on the maximum number of 2's per column. Our approximation scheme is hence an improvement to the
$2^{k-1}$-approximation algorithm except in cases where the maximum number of 2's per row is exponentially small
compared to the maximum number of 2's per column. Our approximation scheme yields also the first approximation results
for $MPPH$.

\medskip

\noindent As explained by Sharan et al. in their ``islands of tractability'' paper \cite{islands}, identifying
tractable special classes can be practically useful for constructing high-speed subroutines within ILP solvers, but
perhaps the most significant aspect of this paper is the analysis underpinning the results, which - by deepening our
understanding of how this problem behaves - assists the search for better, faster approximation algorithms and for
determining the exact shorelines of the islands of tractability.

Furthermore, the fact that - prior to this paper - concrete and positive results for $MPPH$ had not been
obtained (except for rather pessimistic modifications to ILP models \cite{brown}), means that the algorithms given
here for the $MPPH$ cases, and the
 data structures used in their analysis (e.g. the \emph{restricted compatibility graph} in
Section~\ref{sec:posres}), assume particular importance.

Finally, this paper yields some interesting open problems, of which the outstanding $(*,2)$ case (for both
$PH$ and $MPPH$) is only one; prominent amongst these questions (which are discussed at the end of the paper) is the
question of whether $MPPH$ and $PH$ instances are inter-reducible, at least within the bounded-instance framework.

\medskip

\noindent The paper is organised as follows. In Section~\ref{sec:negres} we give the hardness results, in
Section~\ref{sec:posres} we present the polynomial-time solvable cases, in Section~\ref{sec:approx} we give
approximation algorithms and we finish in Section~\ref{sec:concl} with conclusions and open problems.
\section{Hard problems}
\label{sec:negres}
\begin{theorem}
\label{lem:33phyloAPX} $MPPH(3,3)$ is APX-hard.
\end{theorem}
\begin{proof}
The proof in \cite{nphardnote} that $MPPH$ is NP-hard uses a reduction from {\sc Vertex Cover}, which can be modified
to yield NP-hardness and APX-hardness for (3,3)-bounded instances. Given a graph $T=(V,E)$ the reduction in
\cite{nphardnote} constructs a genotype matrix $G(T)$ of $MPPH$ with $|V|+|E|$ rows and $2|V|+|E|$ columns. For every
vertex $v_i \in V$ there is a genotype (row) $g_i$ in $G(T)$ with $g_i(i)=1$, $g_i(i+|V|)=1$ and $g_i(j)=0$ for every
other position $j$. In addition, for every edge $e_k=\{v_{h},v_{l}\}$ there is a genotype $g_k$ with $g_k(h)=2$,
$g_k(l)=2$, $g_k(2|V|+k)=2$ and $g_k(j)=0$ for every other position $j$. Bafna et al. \cite{nphardnote} prove that an
optimal solution for $MPPH$ with input $G(T)$ contains $|V| + |E| + VC(T)$ haplotypes, where $VC(T)$ is the size of
the smallest vertex cover in $T$.

{\sc 3-Vertex Cover} is the vertex cover problem when every vertex in the input graph has at most degree 3. It is
known to be APX-hard \cite{deg3}\cite{cubic}. Let $T$ be an instance of {\sc 3-Vertex Cover}. We assume that $T$ is
connected. Observe that for such a $T$ the reduction described above yields a $MPPH$ instance $G(T)$ that is
$(3,3)$-bounded. We show that existence of a polynomial-time $(1+\epsilon)$ approximation algorithm $A(\epsilon)$ for
$MPPH$ would imply a polynomial-time $(1+\epsilon')$ approximation algorithm for {\sc 3-Vertex Cover} with
$\epsilon'=8\epsilon$.\footnote[1]{Strictly speaking this is insufficient to prove APX-hardness but it is not
difficult to show that the described reduction is actually an L-reduction \cite{deg3}, from which APX-hardness
follows.}

Let $t$ be the solution value for $MPPH(G(T))$ returned by $A(\epsilon)$, and $t^*$ the optimal value for
$MPPH(G(T))$. By the argument mentioned above from \cite{nphardnote} we obtain a solution with value $d = t - |V| -
|E|$ as an approximation of $VC(T)$. Since $t \leq (1+\epsilon)t^*$, we have $d \leq VC(T) + \epsilon VC(T) + \epsilon
|V| + \epsilon |E|$. Connectedness of $T$ implies that $|V|-1 \leq |E|$. In {\sc 3-Vertex Cover}, a single vertex can
cover at most 3 edges in $T$, implying that $VC(T) \geq |E|/3 \geq (|V|-1)/3$. Hence, $|V| \leq 4 VC(T)$ (for $|V|\geq
2$) and we have (if $|V| \geq 2$):
\begin{align*}
d & \leq VC(T) + \epsilon VC(T) + 4\epsilon VC(T) + 3\epsilon VC(T)\\
& \leq VC(T) + 8 \epsilon VC(T)\\
& \leq (1 + 8\epsilon) VC(T).
\end{align*}
\end{proof}
\begin{theorem}
\label{lem:33apx} $PH(3,3)$ is APX-hard.
\end{theorem}
\begin{proof}
\looseness=-1
The proof by Sharan et al. \cite{islands} that $PH(4,3)$ is APX-hard can be modified slightly to obtain
APX-hardness of $PH(3,3)$. The reduction is from {\sc 3-Dimensional Matching} with each element occurring in at most
three triples (3DM3): given disjoint sets $X$, $Y$ and $Z$ containing $\nu$ elements each and a set
$C=\{c_0,\ldots,c_{\mu-1}\}$ of $\mu$ triples in $X\times Y\times Z$ such that each element occurs in at most three
triples in $C$, find a maximum cardinality set $C' \subseteq C$ of disjoint triples.

From an instance of 3DM3 we build a genotype matrix $G$ with $3 \nu
+ 3\mu$ rows and $6\nu+4\mu$ columns. The first $3\nu$ rows are
called \emph{element-genotypes} and the last $3\mu$ rows are called
\emph{matching-genotypes}. We specify non-zero entries of the
genotypes only.\footnote[2]{Only in this proof we index haplotypes,
genotypes and matrices starting with 0, which makes notation
consistent with \cite{islands}.} For every element $x_i \in X$
define element-genotype $g^x_i$ with $g^x_i(3\nu+i)=1$;
$g^x_i(6\nu+4k)=2$ for all $k$ with $x_i \in c_k$. If $x_i$ occurs
in at most two triples we set $g^x_i(i)=2$. For every element $y_i
\in Y$ there is an element-genotype $g^y_i$ with $g^y_i(4\nu+i)=1$;
$g^y_i(6\nu+4k)=2$ for all $k$ with $y_i \in c_k$ and if $y_i$
occurs in at most two triples then we set $g^y_i(\nu + i)=2$. For
every element $z_i \in Z$ there is an element-genotype $g^z_i$ with
$g^z_i(5\nu+i)=1$; $g^z_i(6\nu+4k)=2$ for all $k$ with $z_i \in c_k$
and if $z_i$ occurs in at most two triples then we set
$g^z_i(2\nu+i)=2$. For each triple $c_k=\{ x_{i_1},
y_{i_2},z_{i_3}\} \in C$ there are three matching-genotypes $c_k^x$,
$c_k^y$ and $c_k^z$: $c_k^x$ has $c_k^x(3\nu+i_1)=2$,
$c_k^x(6\nu+4k)=1$ and $c_k^x(6\nu+4k+1)=2$; $c_k^y$ has
$c_k^y(4\nu+i_2)=2$, $c_k^y(6\nu+4k)=1$ and $c_k^y(6\nu+4k+2)=2$;
$c_k^z$ has $c_k^z(5\nu+i_3)=2$, $c_k^z(6\nu+4k)=1$ and
$c_k^z(6\nu+4k+3)=2$.

Notice that the element-genotypes only have a 2 in the first $3\nu$ columns if the element occurs in at most two
triples. This is the only difference with the reduction from \cite{islands}, where every element-genotype has a 2 in
the first $3\nu$ columns: i.e., for elements $x_i\in X$, $y_i\in Y$ or $z_i\in Z$ a 2 in column $i$, $\nu+i$ or
$2\nu+i$, respectively. As a direct consequence our genotype matrix has only three 2's per row in contrast to the four
2's per row in the original reduction.

We claim that for this (3,3)-bounded instance exactly the same arguments can be used as for the (4,3)-bounded
instance. In the original reduction the left-most 2's ensured that, for each element-genotype, at most one of the two
haplotypes used to resolve it was used in the resolution of other genotypes. Clearly this remains true in our modified
reduction for elements appearing in two or fewer triples, because the corresponding left-most 2's have been retained.
So consider an element $x_i$ appearing in three triples and suppose, by way of contradiction, that \emph{both}
haplotypes used to resolve $g^x_i$ are used in the resolution of other genotypes. Now, the 1 in position $3\nu+i$
prevents this element-genotype from sharing haplotypes with other element-genotypes, so genotype $g^x_i$ must share
both its haplotypes with matching-genotypes. Note that, because $g^x_i(3\nu+i)=1$, the genotype $g^x_i$ can only
possibly share haplotypes with matching-genotypes corresponding to triples that contain $x_i$. Indeed, if $x_i$ is in
triples $c_{k_1}$, $c_{k_2}$ and $c_{k_3}$ then the only genotypes with which $g^x_i$ can potentially share haplotypes
are $c^x_{k_1}$, $c^x_{k_2}$ and $c^x_{k_3}$. Genotype $g^x_i$ cannot share both its haplotypes with the same
matching-genotype (e.g. $c^{x}_{k_1}$) because both haplotypes of $g^x_i$ will have a 1 in column $3\nu +i$ whilst
only one of the two haplotypes for $c^{x}_{k_1}$ will have a 1 in that column. So, without loss of generality, $g^x_i$
is resolved by a haplotype that $c^x_{k_1}$ uses and a haplotype that $c^x_{k_2}$ uses. However, this is not possible,
because $g^x_i$ has a 2 in the column corresponding to $c_{k_3}$, whilst both $c^{x}_{k_1}$ and $c^{x}_{k_2}$ have a 0
in that column, yielding a contradiction.

Note that, in the original reduction, it was not only true that each element-genotype shared at most one of its
haplotypes, but - more strongly - it was also true that such a shared haplotype was used by exactly one other genotype
(i.e. the genotype corresponding to the triple the element gets assigned to). To see that this property is also
retained in the modified reduction observe that if (say) $g^x_i$ shares one haplotype with two genotypes $c^{x}_{k_1}$
and $c^{x}_{k_2}$ then $x_i$ must be in both triples $c_{k_1}$ and $c_{k_2}$, but this is not possible because, in the
two columns corresponding to triples $c_{k_1}$ and $c_{k_2}$, $c^{x}_{k_1}$ has 1 and 0 whilst $c^{x}_{k_2}$ has 0 and
1.\\
\end{proof}

\section{Polynomial-time solvability}
\label{sec:posres}
\subsection{Parsimony haplotyping}

\noindent We will prove polynomial-time solvability of $PH$ on (*,1)-bounded instances.

We say that two genotypes $g_1$ and $g_2$ are \emph{compatible}, denoted as $g_1 \sim g_2$, if $g_1(j) =
g_2(j)$ or $g_1(j) = 2$ or $g_2(j) = 2$ for all $j$. A genotype $g$ and a haplotype $h$ are \emph{consistent} if $h$
can be used to resolve $g$, ie. if $g(j)=h(j)$ or $g(j)=2$ for all $j$. The \emph{compatibility graph} is the graph
with vertices for the genotypes and an edge between two genotypes if they are compatible.

\medskip

\begin{figure}
\vspace{-24pt}
\begin{minipage}{.45\textwidth}
\begin{center}
\begin{tabular}{ll}
$\begin{array}{c}
g_1\\
g_2\\
g_3\\
g_4\\
g_5\\
g_6\\
g_7\\
\end{array}$
& $\begin{bmatrix}
0 & 0 & 1 & 0 & 2 & 0 & 1\\
2 & 0 & 2 & 0 & 0 & 0 & 1\\
0 & 0 & 1 & 2 & 0 & 0 & 1\\
0 & 0 & 1 & 0 & 0 & 0 & 2\\
0 & 0 & 1 & 1 & 0 & 2 & 1\\
1 & 2 & 0 & 0 & 0 & 0 & 1\\
0 & 0 & 1 & 1 & 0 & 0 & 1\\
\end{bmatrix}$
\end{tabular}
\end{center}
\end{minipage}
\begin{minipage}{.45\textwidth}
\begin{center}
\epsfig{file=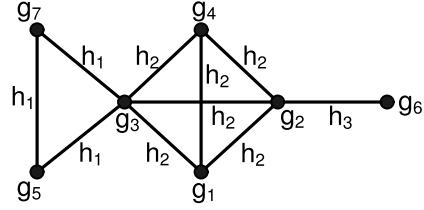} \end{center}
\end{minipage}
\caption{Example of a genotype matrix and the corresponding compatibility graph, with $h_1=(0,0,1,1,0,0,1)$,
$h_2=(0,0,1,0,0,0,1)$ and $h_3=(1,0,0,0,0,0,1)$.} \label{fig:compgraph} \vspace{-12pt}
\end{figure}

\begin{lemma} \label{lem:labelling} If $g_1$ and $g_2$ are compatible rows of a genotype matrix with at most one $2$ per column
then there exists exactly one haplotype that is consistent with both $g_1$ and
$g_2$.\end{lemma}
\begin{proof}
The only haplotype that is consistent with both $g_1$ and $g_2$ is $h$ with $h(j) = g_1(j)$ for all $j$ with $g_1(j)
\neq 2$ and $h(j) = g_2(j)$ for all $j$ with $g_2(j) \neq 2$. There are no columns where $g_1$ and $g_2$ are both
equal to $2$ because there is at most one $2$ per column. In columns where $g_1$ and $g_2$ are both not equal to $2$
they are equal because $g_1$ and $g_2$ are compatible.\\
\end{proof}
\medskip
We use the notation $g_1 \sim_h g_2$ if $g_1$ and $g_2$ are compatible and $h$ is consistent with both. We prove that
the compatibility graph has a specific structure. A \emph{1-sum} of two graphs is the result of identifying a vertex
of one graph with a vertex of the other graph. A 1-sum of $n+1$ graphs is the result of identifying a vertex of a
graph with a vertex of a 1-sum of $n$ graphs. See Figure~\ref{fig:compgraph} for an example of a 1-sum of three
cliques ($K_3$, $K_4$ and $K_2$).

\medskip

\begin{lemma} \label{lem:1sum} If $G$ is a genotype matrix with at most one $2$ per column then every connected component
of the compatibility graph of $G$ is a 1-sum of cliques, where edges in the same clique are labelled with the same
haplotype.
\end{lemma}
\begin{proof}
Let $C$ be the compatibility graph of $G$ and let $g_1,g_2,\ldots,g_k$ be a cycle in $C$. It suffices to show that
there exists a haplotype $h_c$ such that $g_{i} \sim_{h_c} g_{i'}$ for all $i,i'\in\{1,...,k\}$. Consider an arbitrary
column $j$. If there is no genotype with a $2$ in this column then $g_1 \sim g_2 \sim \ldots \sim g_k$ implies that
$g_1(j)=g_2(j)=\ldots = g_k(j)$. Otherwise, let $g_{i_j}$ be the unique genotype with a $2$ in column $j$. Then $g_1
\sim g_2 \sim \ldots \sim g_{i_j-1}$ together with $g_1 \sim g_k \sim g_{k-1}\sim \ldots \sim g_{i_j+1}$ implies that
$g_{i}(j)=g_{i'}(j)$ for all $i,i' \in \{1,...,k\} \setminus \{i_j\}$. Set $h_c(j)=g_i(j)$, $i \neq i_j$. Repeating
this for each column $j$ produces a haplotype $h_c$ such that indeed $g_{i} \sim_{h_c} g_{i'}$ for all
$i,i'\in\{1,...,k\}$.\\
\end{proof}

\medskip
From this lemma, it follows directly that in $PH(*,1)$ the compatibility graph is {\em chordal}, meaning
that all its induced cycles are triangles. Every chordal graph has a \emph{simplicial} vertex, a vertex whose (closed)
neighbourhood is a clique. Deleting a vertex in a chordal graph gives again a chordal graph (see for example
\cite{blair} for an introduction to chordal graphs). The following lemma leads almost immediately to polynomial
solvability of $PH(*,1)$. We use set-operations for the rows of matrices: thus, e.g., $h\in H$ says $h$ is a row of
matrix $H$, $H\cup h$ says $h$ is added to $H$ as a row, and $H'\subset H$ says $H'$ is a submatrix consisting of rows
of $H$.

\medskip

\begin{lemma} \label{lem:starone} Given haplotype matrix $H'$ and genotype
matrix $G$ with at most one 2 per column it is possible to find, in polynomial time, a haplotype matrix $H$ that
resolves $G$, has $H'$ as a submatrix and has a minimum number of rows.
\end{lemma}
\begin{proof}
\looseness=-1 The proof is constructive. Let problem $(G,H')$ denote the above problem on input matrices $G$ and $H'$.
Let $C$ be the compatibility graph of $G$, which implied by Lemma~\ref{lem:1sum} is chordal. Suppose $g$ corresponds
to a simplicial vertex of $C$. Let $h_c$ be the unique haplotype consistent with any genotype in the closed
neighbourhood clique of $g$. We extend matrix $H'$ to $H''$ and update graph $C$ as follows.
\begin{enumerate}
\item If $g$ has no $2$'s it can be resolved with only one haplotype $h=g$. We set $H''=H'\cup h$ and remove $g$ from $C$.
\item Else, if there exist rows $h_1\in H'$ and $h_2\in H'$ that resolve $g$ we set $H''=H'$ and remove $g$ from $C$.
\item Else, if there exists $h_1\in H'$ such that $g=h_1+h_c$ we set $H''=H'\cup h_c$ and remove $g$ from $C$.
\item Else, if there exists $h_1\in H'$ and $h_2\notin H'$ such that $g=h_1+h_2$ we set $H''=H'\cup h_2$ and remove $g$ from $C$.
\item Else, if $g$ is not an isolated vertex in $C$ then there exists a haplotype $h_1$ such that $g=h_1+h_c$ and we set
$H''=H'\cup \{h_1, h_c\}$ and remove $g$ from $C$.
\item Otherwise, $g$ is an isolated vertex in $C$ and we set $H''=H'\cup \{h_1, h_2\}$ for any $h_1$ and $h_2$ such that
$g=h_1+h_2$ and remove $g$ from $C$.
\end{enumerate}
The resulting graph is again chordal and we repeat the above procedure for $H'=H''$ until all vertices are removed from $C$.
Let $H$ be the final haplotype matrix $H''$. It is clear from the construction that $H$ resolves $G$.

We prove that $H$ has a minimum number of rows by induction on the number of genotypes. Clearly, if $G$ has only one
genotype the algorithm constructs the only, and hence optimal, solution. The induction hypothesis is that the
algorithm finds an optimal solution to the problem $(G,H')$ for any haplotype matrix $H'$ if $G$ has at most $n-1$
rows. Now consider haplotype matrix $H'$ and genotype matrix $G$ with $n$ rows. The first step of the algorithm
selects a simplicial vertex $g$ and proceeds with one of the cases 1 to 6. The algorithm then finds (by the induction
hypothesis) an optimal solution $H$ to problem $(G\setminus\{g\},H'')$. It remains to prove that $H$ is also an
optimal solution to problem $(G,H')$. We do this by showing that an optimal solution $H^*$ to problem $(G,H')$ can be
modified  to include $H''$. We prove this for every case of the algorithm separately.

\begin{enumerate}
\item In this case $h\in H^*$, since $g$ can only be resolved by $h$.\smallskip \item In this case $H''=H'$ and hence
$H''\subseteq H^*$.\smallskip \item Suppose that $h_c \notin H^*$. Because we are not in case $2$ we know that there
are two rows in $H^*$ that resolve $g$ and at least one of the two, say $h^*$, is not a row of $H'$. Since $h_c$ is
the unique haplotype consistent with (the simplicial) $g$ and any compatible genotype, $h^*$ can not be consistent
with any other genotype than $g$. Thus, replacing $h^*$ by $h_c$ gives a solution with the same number of rows but
containing $h_c$. \smallskip \item Suppose that $h_2\notin H^*$. Because we are not in case $2$ or $3$ we know that
there is a haplotype $h^*\in H^*$ consistent with $g$, $h^*\notin H'$ and $h^*\neq h_c$. Hence it is not consistent
with any other genotypes than $g$ and we can replace $h^*$ by $h_2$. \smallskip \item Suppose that $h_1\notin H^*$ or
$h_c\notin H^*$. Because we are not in case $2$, $3$ or $4$, there are haplotypes $h^*\in H\backslash H'$ and
$h^{**}\in H\backslash H'$ that resolve $g$. If $h^*$ and $h^{**}$ are both not equal to $h_c$ then they are not
consistent with any other genotype than $g$. Replacing $h^*$ and $h^{**}$ by $h_1$ and $h_c$ leads to another optimal
solution. If one of $h^*$ and $h^{**}$ is equal to $h_c$ then we can replace the other one by $h_1$. \smallskip \item
\looseness=-1 Suppose that $h_1\notin H^*$ or $h_2\notin H^*$. There are haplotypes $h^*,h^{**}\in H^*\backslash H'$
that resolve $g$ and just $g$ since $g$ is an isolated vertex. Replacing $h^*$ and $h^{**}$ by $h_1$ and $h_2$ gives
an optimal solution containing $h_1$ and $h_2$.
\end{enumerate}
\end{proof}
\begin{theorem} \label{prop:starone}
The problem $PH(*,1)$ can be solved in polynomial time. \end{theorem}
\begin{proof}
The proof follows from Lemma~\ref{lem:starone}. Construction of the compatibility graph takes $O(n^2m)$ time, for an
$n$ times $m$ input matrix. Finding an ordering in which to delete the simplicial vertices can be done in time
$O(n^2)$ \cite{rose} and resolving each vertex takes $O(n^2m)$ time. The overall running time of the algorithm is
therefore $O(n^3m)$.\\
\end{proof}

\subsection{Minimum perfect phylogeny haplotyping}

\noindent
\looseness=-1 Polynomial-time solvability of $PH$ on $(2,*)$-bounded instances has been shown in \cite{wabi} and
\cite{lancia}. We prove it for $MPPH(2,*)$. We start with a definition.

\medskip

\begin{definition} \label{def:redres} For two columns of a genotype matrix we say that a \emph{reduced resolution} of these columns
is the result of applying the following rules as often as possible to the submatrix induced by these columns: deleting
one of two identical rows and the replacement rules\\ $\begin{bmatrix} 2 & a \end{bmatrix} \rightarrow
\begin{bmatrix} 1 & a \\ 0 & a \end{bmatrix}$, $\begin{bmatrix} a & 2 \end{bmatrix} \rightarrow \begin{bmatrix} a & 1
\\ a & 0 \end{bmatrix}$, $\begin{bmatrix} 2 & 2 \end{bmatrix} \rightarrow \begin{bmatrix} 1 & 1 \\ 0 & 0 \end{bmatrix}$ and
$\begin{bmatrix} 2 & 2 \end{bmatrix} \rightarrow \begin{bmatrix} 1 & 0 \\ 0 & 1 \end{bmatrix}$, for $a \in \{0,1\}$.\\
\end{definition}
Note that two columns can have more than one reduced resolution if there is a genotype with a 2 in both these columns.
The reduced resolutions of a column pair of a genotype matrix $G$ are submatrices of (or equal to) $F$ and represent
all possibilities for the submatrix induced by the corresponding two columns of a minimal haplotype matrix $H$
resolving $G$, after collapsing identical rows.

\medskip

\begin{theorem} \label{prop:startwophylo}
The problem $MPPH(2,*)$ can be solved in polynomial time.
\end{theorem}
\begin{proof}
We reduce $MPPH(2,*)$ to $PH$(2,*), which can be solved in polynomial time (see above). Let $G$ be an instance of
$MPPH(2,*)$. We may assume that any two rows are different.

Take the submatrix of any two columns of $G$. If it does not contain a [2~2] row, then in terms of
Definition~\ref{def:redres} there is only one reduced resolution. If $G$ contains two or more [2~2] rows then,
since by assumption all genotypes are different, $G$ must have $\begin{bmatrix} 2 & 2 & 0 \\
2 & 2 & 1 \end{bmatrix}$ and therefore $\begin{bmatrix} 2 & 0 \\
2 & 1 \end{bmatrix}$ as a submatrix, which can only be resolved by a haplotype matrix containing the forbidden
submatrix $F$. It follows that in this case the instance is infeasible. If it contains exactly one [2~2] row, then
there are clearly two reduced resolutions. Thus we may assume that for each column pair there are at most two reduced
solutions.

Observe that if for some column pair all reduced resolutions are equal to $F$ the instance is again infeasible. On the
other hand, if for all column pairs none of the reduced resolutions is equal to $F$ then $MPPH(2,*)$ is equivalent to
$PH(2,*)$ because any minimal haplotype matrix $H$ that resolves $G$ admits a perfect phylogeny. Finally, consider a
column pair with two reduced resolutions, one of them containing $F$. Because there are two reduced resolutions there
is a genotype $g$ with a 2 in both columns. Let $h_1$ and $h_2$ be the haplotypes that correspond to the resolution of
$g$ that does not lead to $F$. Then we replace $g$ in $G$ by $h_1$ and $h_2$, ensuring that a minimal haplotype matrix
$H$ resolving $G$ can not have $F$ as a submatrix in these two columns.

Repeating this procedure for every column pair either tells us that the matrix $G$ was an infeasible instance or
creates a genotype matrix $G'$ such that any minimal haplotype matrix $H$ resolves $G'$ if and only if $H$ resolves
$G$, and $H$ admits a perfect phylogeny.\\
\end{proof}

\medskip

\begin{theorem} \label{prop:staronephylo} The problem $MPPH(*,1)$ can be solved in polynomial time.
\end{theorem}
\begin{proof}
Similar to the proof of Theorem~\ref{prop:startwophylo} we reduce $MPPH(*,1)$ to $PH(*,1)$. As there, consider for any
pair of columns of the input genotype matrix $G$ its reduced resolutions, according to Definition~\ref{def:redres}. Since
$G$ has at most one $2$ per column there is at most one genotype with 2's in both columns. Hence there are at most two
reduced resolutions. If all reduced resolutions are equal to the forbidden submatrix $F$ the instance is infeasible.
If on the other hand for all column pairs no reduced resolution is equal to $F$ then in fact $MPPH(*,1)$ is equivalent
to $PH(*,1)$, because any minimal haplotype matrix resolving $G$ admits a perfect phylogeny.

As in the proof of Theorem~\ref{prop:startwophylo} we are left with considering column pairs for which one of the two
reduced resolutions is equal to $F$. For such a column pair there must be a genotype $g$ that has 2's in both these
columns. The other genotypes have only 0's and 1's in them. Suppose we get a forbidden submatrix $F$ in these columns
of the solution if $g$ is resolved by haplotypes $h_1$ and $h_2$, where $h_1$ has $a$ and $b$ and therefore $h_2$ has
$1-a$ and $1-b$ in these columns, $a,b\in \{0,1\}$. We will change the input matrix $G$ such that if $g$ gets resolved
by such a \emph{forbidden resolution} these haplotypes are not consistent with any other genotypes. We do this by
adding an extra column to $G$ as follows. The genotype $g$ gets a $1$ in this new column. Every genotype with $a$ and
$b$ or with $1-a$ and $1-b$ in the considered columns gets a $0$ in the new column. Every other genotype gets a $1$ in
the new column. For example, the matrix
\[
\begin{bmatrix} 2 & 2 \\ 0 & 1 \\ 1 & 0 \\ 1 & 1 \end{bmatrix}
{\rm \ gets\ one\ extra\ column\ and\ becomes}
\begin{bmatrix} 2 & 2 & 1 \\ 0 & 1 & 1\\ 1 & 0 & 1\\ 1 & 1 & 0\end{bmatrix}.
\]
\noindent Denote by  $G_{mod}$  the result of modifying $G$ by adding such a column for every pair of columns with
exactly one `bad' and one `good' reduced resolution. It is not hard to see that any optimal solution to $PH(*,1)$ on
$G_{mod}$ can be transformed into a solution to $MPPH(*,1)$ on $G$ of the same cardinality (indeed, any two haplotypes
used in a forbidden resolution of a genotype $g$ in $G_{mod}$ are not consistent with any other genotype of $G_{mod}$,
and hence may be replaced by two other haplotypes resolving $g$ in a non-forbidden way). Now, let $H$ be an optimal
solution to $MPPH(*,1)$ on $G$. We can modify $H$ to obtain a solution to $PH(*,1)$ on $G_{mod}$ of the same
cardinality as follows. We modify every haplotype in $H$ in the same way as the genotypes it resolves. From the
construction of $G_{mod}$ it follows that two compatible genotypes are only modified differently if the haplotype they
are both consistent with is in a forbidden resolution. However, in $H$ no genotypes are resolved with a forbidden
resolution since $H$ is a solution to $MPPH(*,1)$. We conclude that optimal solutions to $PH(*,1)$ on $G_{mod}$
correspond to optimal solutions to $MPPH(*,1)$ on $G$ and hence the latter problem can be solved in polynomial time,
by Theorem \ref{prop:starone}.

If we use the algorithm from the proof of Lemma~\ref{lem:starone} as a subroutine we get an overall running time of
$O(n^3m^2)$, for an $n \times m$ input matrix.\\
\end{proof}
\medskip
\medskip
The borderline open complexity problems are now $PH(*,2)$ and $MPPH(*,2)$. Unfortunately, we have not found the answer
to these complexity questions. However, the borders have been pushed slightly further. In \cite{islands} $PH(*,2)$ is
shown to be polynomially solvable if the input genotypes have the complete graph as compatibility graph, we call this
problem $PH(*,2)$-$C1$. We will give the counterpart result for $MPPH(*,2)$-$C1$.

Let $G$ be an $n \times m$ $MPPH(*,2)$-$C1$ input matrix. Since the compatibility graph is a clique, every column of
$G$ contains only one symbol besides possible 2's. If we replace in every 1-column of $G$ (a column containing only
1's and 2's) the 1's by 0's and mark the SNP corresponding to this column `flipped', then
 we obtain an equivalent problem
on a $\{0,2\}$-matrix $G'$.
To see that this problem is indeed equivalent, suppose $H'$ is a haplotype matrix
resolving this modified genotype
matrix $G'$ and suppose $H'$ does not contain the forbidden submatrix $F$.
 Then by interchanging 0's and 1's in every column of $H'$
corresponding to a flipped SNP, one obtains a haplotype matrix $H$ without the forbidden submatrix
which resolves the original input matrix $G$. And vice versa.
Hence, from now on we will assume, without loss of generality, that the input matrix $G$ is a $\{0,2\}$-matrix.

If we assume moreover that $n\geq 3$, which we do from here on, the \emph{trivial haplotype} $h_t$ defined as the
all-0 haplotype of length $m$ is the only haplotype consistent with all genotypes in $G$.

We define the \emph{restricted} compatibility graph $C_{R}(G)$ of
$G$ as follows. As in the normal compatibility graph, the vertices of
$C_{R}(G)$ are the genotypes of $G$. However, there is an edge
$\{g,g'\}$ in $C_{R}$(G) only if $g \sim_{h} g'$ for some $h \neq
h_t$, or, equivalently, if there is a column where both $g$ and $g'$
have a 2.

\medskip

\begin{lemma}
\label{lem:deg2} If $G$ is a feasible instance of $MPPH(*,2)$-$C1$
then every vertex in $C_R(G)$ has degree at most 2.
\end{lemma}
\begin{proof}
Any vertex of degree higher than 2 in $C_R(G)$ implies the existence
in $G$ of submatrix:

\medskip
\[
B= \begin{bmatrix} 2 & 2 & 2 \\ 2 & 0 & 0 \\ 0 & 2 & 0 \\ 0 & 0 & 2 \end{bmatrix}
\]
\medskip

\noindent \looseness=+1 It is easy to verify that no resolution of this submatrix permits a perfect phylogeny.\\
\end{proof}

\medskip

Suppose that $G$ has two identical columns. There are either 0, 1 or 2 rows with 2's in both these columns.
In each case it is easy to see that any haplotype matrix $H$ resolving $G$ can be modified, without introducing
 a forbidden submatrix,  to make
the corresponding columns in $H$ equal as well (simply delete one column and duplicate another). This leads to the
first step of the algorithm {\bf A} that we propose for solving $MPPH(*,2)$-$C1$:

\medskip

\noindent {\bf Step 1 of A}: Collapse all identical columns in $G$.

\medskip

\noindent From now on,
 we assume that there are no identical columns. Let us partition the genotypes in $G_0$, $G_1$
and $G_2$, denoting the set of genotypes in $G$ with, respectively, degree 0,1, and 2 in $C_R(G)$. For any genotype
$g$ of degree 1 in $C_R(G)$ there is exactly
 one genotype with a 2 in the same column as $g$. Because there are no
identical columns,
 it follows that any genotype $g$ of degree 1 in $C_R(G)$ can have at most two 2's. Similarly any
genotype of degree 2 in $C_R(G)$ has at most three 2's. Accordingly we define $G_1^1$ and $G_1^2$ as the genotypes in
$G_1$ that have one 2 and two 2's, respectively, and similarly $G_2^2$ and $G_2^3$ as the genotypes in $G_2$ with two
and three 2's, respectively.

The following lemma states how genotypes in these sets  must  be resolved if no submatrix $F$ is allowed in
the solution. If genotype $g$ has $k$ 2's we denote by  $g[a_1,a_2,\ldots,a_k]$ the haplotype
 with entry $a_i$ in the position  where $g$ has its $i$-th 2  and 0 everywhere else.

\medskip

\begin{lemma}
\label{lem:usegeno} A haplotype matrix is a feasible solution to the problem $MPPH(*,2)$-$C1$ if and only if all genotypes are resolved in one of the following ways:

\noindent {\em (i)} A genotype $g\in G_1^1$ is resolved by $g[1]$ and $g[0]=h_t$. \\
{\em (ii)} A genotype $g\in G_2^2$ is resolved by $g[0,1]$ and $g[1,0]$. \\
{\em (iii)} A genotype $g\in G_1^2$ is either resolved by $g[0,0]=h_t$
and $g[1,1]$ or by $g[0,1]$ and $g[1,0]$. \\
{\em (iv)} A genotype $g\in G_2^3$ is either resolved by $g[1,0,0]$
and $g[0,1,1]$ or by $g[0,1,0]$ and $g[1,0,1]$ (assuming that
 the two neighbours of $g$ have a 2 in the first two positions where $g$ has a 2).
\end{lemma}
\begin{proof}
A genotype $g\in G_2^2$ has degree 2 in $C_R(G)$, which implies the existence in $G$ of a submatrix:
\medskip
\begin{center}
$D =$
\begin{tabular}{ll}
$\begin{array}{l}
g\\
g'\\
g''\\
\end{array}
\begin{bmatrix} 2 & 2 \\ 2 & 0 \\ 0 & 2 \end{bmatrix}$
\end{tabular}.
\end{center}
\medskip
\noindent Resolving $g$ with $g[0,0]$ and $g[1,1]$ clearly leads to the forbidden submatrix $F$. Similarly, resolving
a genotype $g\in G_2^3$ with $g[0,0,1]$ and $g[1,1,0]$ or with $g[0,0,0]$ and $g[1,1,1]$ leads to a forbidden
submatrix in the first two columns where $g$ has a 2. It follows that
 resolving the genotypes in a way other than
 described in the lemma
yields a haplotype matrix which does not admit a perfect phylogeny.

Now suppose that all genotypes are resolved as described in the lemma and assume that there is a forbidden submatrix
$F$ in the solution. Without loss of generality,  we assume $F$ can be found in the first two columns of the solution
matrix. We may also assume that no haplotype can be deleted from the solution. Then, since $F$ contains [1 1], there
is a genotype $g$ starting with [2~2]. Since there are no identical columns there are only two possibilities. The
first possibility is that there is exactly one other genotype $g'$ with a 2 in exactly one of the first two columns.
Since all genotypes different from $g$ and $g'$ start with [0 0], none of the resolutions of $g$ can have created the
complete submatrix $F$. Contradiction. The other possibility is that there is exactly one genotype with a 2 in the
first column and exactly one genotype with a 2 in the second column, but these are different genotypes, i.e. we have
the submatrix $D$. Then $g\in G_2^3$ or $g\in G_2^2$ and it can again be checked that none of the resolutions in (ii)
and (iv) leads to the forbidden submatrix.\\
\end{proof}

\medskip

\begin{lemma} Let $G$ be an instance of $MPPH(*,2)$ and $G_1^2$, $G_2^3$ as defined above.
\label{lem:private} \\ {\em (i)} Any nontrivial haplotype is consistent
with at most two genotypes in $G$.\\
{\em (ii)} A genotype $g\in G_1^2\cup G_2^3$ must be resolved using at least one haplotype that is not consistent with
any other genotype.
\end{lemma}
\begin{proof} {\em (i)} Let $h$ be a nontrivial haplotype.
There is a column where $h$ has a 1 and there are at most
two genotypes with a 2 in that column. \\
{\em (ii)} A genotype $g\in G_1^2\cup G_2^3$ has a 2 in a column that has no other 2's. Hence there is a haplotype
with a 1 in this column and this haplotype is not consistent with any other genotypes.\\
\end{proof}

\medskip

 A haplotype that is only consistent with $g$ is called a \emph{private haplotype} of $g$. Based on (i) and
(ii) of Lemma~\ref{lem:usegeno} we propose the next step of {\bf A}:

\medskip

\noindent {\bf Step 2 of A}: \looseness=-1 Resolve all $g\in G_1^1 \cup G_2^2$ by the unique haplotypes allowed to
resolve them according to Lemma~\ref{lem:usegeno}. Also resolve each $g\in G_0$ with $h_t$ and the complement of $h_t$
with respect to $g$. This leads to a partial haplotype matrix $H_2^p$.

\medskip

\noindent The next step of {\bf A} is based on Lemma~\ref{lem:private} (ii).

\medskip
\noindent {\bf Step 3 of A}: \looseness=-1 For each $g\in G_1^2 \cup G_2^3$ with $g\sim_{h'}g'$ for some $h'\in H_2^p$
that is allowed to resolve $g$ according to Lemma~\ref{lem:usegeno}, resolve $g$ by adding the complement $h''$ of
$h'$ w.r.t. $g$ to the set of haplotypes, i.e. set $H_2^p := H_2^p \cup \{h''\}$, and repeat this step as long as new
haplotypes get added. This leads to partial haplotype matrix $H_3^p$.

\medskip

\noindent Notice that $H_3^p$ does not contain any haplotype that is
allowed to resolve any of the genotypes that have not been resolved
in Steps 2 and 3. Let us denote this set of leftover, unresolved
haplotypes by $GL$, the degree 1 vertices among those by  $GL_1\subseteq G_1^2$, and the
degree 2 vertices among those  by $GL_2\subseteq G_2^3$. The restricted
compatibility graph induced by $GL$, which we denote by $C_R(GL)$
consists of paths and circuits. We first give the final steps of
algorithm A and argue optimality afterwards.

\medskip

\noindent {\bf Step 4 of A}: Resolve each cycle in $C_R(GL)$, necessarily consisting of $GL_2$-vertices, by starting
with an arbitrary vertex and, following the cycle, resolving each next pair $g,g'$ of vertices by haplotype $h \neq
h_t$ such that $g\sim_h g'$ and the two complements of $h$ w.r.t. $g$ and $g'$ respectively. In case of an odd cycle
the last vertex is resolved by any pair of haplotypes that is allowed to resolve it. Note that $h$ has a 1 in the
column where both $g$ and $g'$ have a 2 and otherwise 0. It follows easily that $g$ and $g'$ are both allowed to use
$h$ (and its complement) according to (iv) of Lemma~\ref{lem:usegeno}.

\medskip

\noindent {\bf Step 5 of A}: Resolve each path in $C_R(GL)$ with both endpoints in $GL_1$ by first resolving the
$GL_1$ endpoints by the trivial haplotype $h_t$ and the complements of $h_t$ w.r.t. the two endpoint genotypes,
respectively. The remaining path contains only $GL_2$-vertices and is resolved according to Step 6.

\medskip

\noindent {\bf Step 6 of A}: Resolve each remaining path by starting in (one of) its $GL_2$-endpoint(s), and following
the path, resolving each next pair of vertices as in Step 4. In case of a path with an odd number of vertices, resolve
the last vertex by any pair of haplotypes that is allowed to resolve it in case it is a $GL_2$-vertex, and resolve it
by the trivial haplotype and its complement w.r.t. the vertex in case it is a $GL_1$ vertex.

\medskip

By construction the haplotype matrix $H$ resulting from {\bf A} resolves $G$. In addition, from
Lemma~\ref{lem:usegeno} follows that $H$ admits a perfect phylogeny.

To argue minimality of the solution, first observe that the haplotypes added in Step 2 and Step 3 are
unavoidable by Lemma~\ref{lem:usegeno} (i) and (ii) and Lemma~\ref{lem:private} (ii). Lemma~\ref{lem:private} tells us
moreover that the resolution of a cycle of $k$ genotypes in $GL_2$ requires at least $k+\lceil\frac{k}{2}\rceil$
haplotypes that can not be used to resolve any other genotypes in $GL$. This proves optimality of Step 4. To prove
optimality of the last two steps we need to take into account that genotypes in $GL_1$ can potentially share the
trivial haplotype. Observe that to resolve a path with $k$ vertices one needs at least $k+\lceil\frac{k}{2}\rceil$
haplotypes. Indeed {\bf A} does not use more than that in Steps 5 and 6. Moreover, since these paths are disjoint,
they cannot share haplotypes for resolving their genotypes except for the endpoints if they are in $GL_1$, which can
share the trivial haplotype. Indeed, {\bf A} exploits the possibility of sharing the trivial haplotype in a maximal
way, except on a path with an even number of vertices and one endpoint in $GL_1$. Such a path, with $k$ (even)
vertices, is resolved in {\bf A} by $3\frac{k}{2}$ haplotypes that can not be used to resolve any other genotypes. The
degree 1 endpoint might alternatively be resolved by the trivial haplotype and its complement w.r.t. the corresponding
genotype, adding the latter private haplotype, but then for resolving the remaining path with $k-1$ (odd) vertices
only from $GL_2$ we still need $k-1+\lceil\frac{k-1}{2}\rceil$, which together with the private haplotype of the
degree 1 vertex gives $3\frac{k}{2}$ haplotypes also (not even counting $h_t$).

As a result we have polynomial-time solvability of $MPPH(*,2)$-$C1$.

\medskip

\begin{theorem}
$MPPH(*,2)$ is solvable in polynomial time if the compatibility graph is a clique.
\flushright
\QEDclosed
\end{theorem}

\section{Approximation algorithms}
\label{sec:approx}
In this section we construct polynomial time approximation algorithms for $PH$ and $MPPH$, where the accuracy depends
on the number of 2's per column of the input matrix. We describe genotypes without 2's as \emph{trivial} genotypes,
since they have to be resolved in a trivial way by one haplotype. Genotypes with at least one 2 will be described as
\emph{nontrivial} genotypes. We write \phminZ{} and \mpphminZ{} to denote the restricted versions of the problems
where each genotype is nontrivial. We make this distinction between the problems because we have better lower bounds
(and thus approximation ratios) for the restricted variants.

\subsection{$PH$ and $MPPH$ where all input genotypes are nontrivial}
To prove approximation guarantees we need good lower bounds on the number of haplotypes in the solution. We start with
two bounds from \cite{islands}, whose proof we give because the first one is short but based on a crucial observation, and the second one was incomplete in \cite{islands}. We use these bounds to obtain a different lower
bound that we need for our approximation algorithms. \medskip
\begin{lemma}
\label{lem:minBound} \cite{islands} Let $G$ be an $n \times m$ instance of \phminZ{} (or \mpphminZ). Then at
least
\begin{eqnarray*}
LB_{sqrt}(n) = \bigg \lceil \frac{ 1 + \sqrt{1+8n} }{2} \bigg \rceil
\end{eqnarray*}
haplotypes are required to resolve $G$.
\end{lemma}
\begin{proof}
The proof follows directly from the observation that $q$ haplotypes can resolve at most $\binom{q}{2} = q(q-1)/2$
nontrivial genotypes.\\
\end{proof}
\medskip
\begin{lemma}
\label{lem:theirbound} \cite{islands} Let $G$ be an $n \times m$ instance of \phmin{\ell}, for some $\ell \geq 1$,
such that the compatibility graph of $G$ is a clique. Then at least
\begin{eqnarray*}
LB_{sha}(n,\ell) = \bigg \lceil \frac{2n}{\ell+1} + 1 \bigg \rceil
\end{eqnarray*}
haplotypes are required to resolve $G$.
\end{lemma}
\begin{proof}
Recall that, after relabeling if necessary, the trivial haplotype $h_t$ is the all-0 haplotype and is consistent with all genotypes. Suppose a solution of $G$ has $q$
non-trivial haplotypes. Observe that $h_t$ can be used in the resolution of at most $q$ genotypes. Also observe (by
Lemma 5 in \cite{islands}) that each non-trivial haplotype can be used in the resolution of at most $\ell$ genotypes.
Now distinguish two cases. First consider the case where $h_t$ is in the solution. Then from the two observations
above it follows that $n \leq (q+\ell q)/2$ and hence the solution consists of at least $q+1 \geq 2n/(\ell +1)+1$
haplotypes. Now consider the second case i.e. where $h_t$ is not in the solution. Then we have that $n \leq \ell q/2$
and hence that the solution consists of at least $2n/\ell$ haplotypes. If $n \geq \ell (\ell+1)/2$ we have that
$2n/\ell \geq 2n/(\ell+1)+1$, and the claim follows. If $n < \ell (\ell+1)/2$ then this implies that $\ell
>\frac{\sqrt{1+8n}-1}{2}$. Combining this with that by Lemma~\ref{lem:minBound} $q\geq \frac{\sqrt{1+8n}+1}{2}$ gives
that $(\ell+1)(q-1) > \frac{1}{4}(\sqrt{1+8n} + 1)(\sqrt{1+8n} - 1)$, which is equal to $2n$. It follows that $q >
2n/(\ell +1)+1$.\\
\end{proof}
\medskip
The $LB_{sha}$ bound has been proven only for \phminZ{} (and \mpphminZ) instances where the compatibility graph is a
clique. We now prove a different bound which, in terms of cliques, is slightly weaker (for large $n$) than $LB_{sha}$,
but which allows us to generalise the bound to more general inputs. (Indeed it remains an open question whether
$LB_{sha}$ applies as a lower bound not just for cliques but also for general instances.)
\medskip
\begin{lemma}
\label{lem:boundmin} Let $G$ be an $n \times m$ instance of \phmin{\ell}, for some $\ell \geq 1$. Then at least
\begin{equation}
\lbmidmin{n}{\ell} = \bigg \lceil \frac{2(n+\ell)(\ell+1)}{\ell(\ell+3)}
\bigg \rceil
\end{equation}
haplotypes are required to resolve $G$.
\end{lemma}
\begin{proof}
Let $C(G)$ be the compatibility graph of $G$. We may assume without loss of generality that $C(G)$ is connected. First
consider the case where $C(G)$ is a clique. If $n \geq \ell ( \ell +1)/2$, it suffices to notice that
$\lbmidmin{n}{\ell}\leq LB_{sha}(n,\ell)$ for each value of $\ell \geq 1$, since the function
\begin{equation}
f(n) = \frac{2n}{\ell +1}+1 - \frac{2(n+\ell)(\ell+1)}{\ell(\ell+3)}
\end{equation}
is equal to $0$ if $n= \ell ( \ell +1)/2$ and has nonnegative derivative
$f'(n)=\frac{2}{\ell+1}-2\frac{\ell+1}{\ell(\ell+3)}\geq 0$.\\
Secondly, if $1 \leq n \leq \ell (\ell +1)/2$, straightforward but tedious calculations show that for all $\ell \geq
1$ the function
\begin{equation}
 F(n)= \frac{ 1 + \sqrt{1+8n}}{2} - \frac{2(n+\ell)(\ell+1)}{\ell(\ell+3)}
\end{equation}
has value $0$ for $n= \ell ( \ell +1)/2$ and for some $n$ in the interval $[0,1]$, whereas in between these values it
has positive value. Hence, $\lbmidmin{n}{\ell}\leq LB_{sqrt}(n)$ for $1 \leq n \leq \ell (\ell +1)/2$.

To prove that the bound also holds if $C(G)$ is not a clique we use induction on $n$. Suppose that
for each $n'< n$ the lemma
holds for all $n' \times m$ instances $G'$ of \phmin{\ell '} for every $m$ and $\ell '$.
Since $C(G)$ is not a clique there exist two genotypes $g_1$ and $g_2$ in $G$ and a
column $j$ such that $g_1(j)=0$ and $g_2(j)=1$. Given that $G$ is a \phmin{\ell} instance
$t \leq \ell$ genotypes have a 2 in column $j$.
Deleting these $t$ genotypes yields an instance $G^d$ with disconnected compatibility graph $C(G^d)$, since the absence of a $2$ in column $j$ prevents the existence of any path from $g_1$ to $g_2$. Let $C(G^d)$ have $p \geq 2$
components $C(G_1), ..., C(G_p)$, and let $n_i \geq 1$ denote the number of genotypes in $G_i$. Thus, $n = n_1 +
... + n_p + t$. We use the induction hypothesis on $G_1,\ldots,G_p$ to conclude that the number of haplotypes required to resolve $G$ is at least
\begin{eqnarray*}
\sum_{i=1}^p \bigg \lceil \frac{2(n_i + \ell)(\ell+1)}{\ell(\ell+3)} \bigg \rceil
              & \geq & \bigg \lceil \frac{2(\sum_{i=1}^p n_i + p\ell)(\ell+1)}{\ell(\ell+3)} \bigg \rceil
             \geq \bigg \lceil \frac{2(\sum_{i=1}^p n_i + 2\ell)(\ell+1)}{\ell(\ell+3)} \bigg \rceil \\
             & \geq & \bigg \lceil \frac{2(\sum_{i=1}^p n_i + t+ \ell)(\ell+1)}{\ell(\ell+3)} \bigg \rceil
             = \bigg \lceil \frac{2(n + \ell)(\ell+1)}{\ell(\ell+3)} \bigg \rceil
\end{eqnarray*}
\end{proof}
\medskip
\begin{corollary}
\label{cor:easyapproxMin} Let $G$ be an $n \times m$ instance of \phmin{\ell} or \mpphmin{\ell}, for some $\ell \geq
1$. Any feasible solution for $G$ is within a ratio $\ell + 2 - \frac{2}{\ell+1}$ from optimal.
\end{corollary}
\begin{proof}
Immediate from the fact that any solution for $G$ has at most $2n$ haplotypes. In the case of $MPPH$ we can check
whether feasible solutions exist, and if so obtain such a solution, by using the algorithm in for example
\cite{gusfieldlinear}.\\
\end{proof}
\medskip
Not surprisingly, better approximation ratios can be achieved. The following simple algorithm computes
approximations of \phmin{\ell}. (The algorithm does not work for $MPPH$, however.)

\medskip

\noindent
\textbf{Algorithm:} $PH^{nt}M$ \\
\textbf{Step 1:} construct the compatibility graph $C(G)$.\\
\textbf{Step 2:} find a maximal matching $M$ in $C(G)$.\\
\textbf{Step 3:} for every edge $\{g_1,g_2\}\in M$, resolve $g_1$ and $g_2$ by in total 3 haplotypes: any haplotype
consistent with both $g_1$ and $g_2$, and its complements with respect to $g_1$ and $g_2$.\\
\textbf{Step 4:} resolve each remaining genotype by two haplotypes.
\medskip
\begin{theorem}
$PH^{nt}M$ computes a solution to \phmin{\ell} in polynomial time within an approximation
ratio of $c(\ell)=\frac{3}{4}\ell +\frac{7}{4}-\frac{3}{2}\frac{1}{\ell +1}$, for every $\ell \geq 1$.
\end{theorem}
\begin{proof}
Since constructing $C(G)$ given $G$ takes $O(n^2m)$ time and finding a maximal matching in any graph takes linear
time, $O(n^2m)$ running time follows directly.

Let $q$ be the size of the maximal matching.
Then $PH^{nt}M$ gives a solution with
$3q+2(n-2q)$ = $2n-q$ haplotypes. Since the complement of the
maximal matching is an independent set of size $n-2q$, any solution must contain at least $2(n-2q)$
haplotypes to resolve the genotypes in this independent set.
The theorem thus holds if $\frac{2n-q}{2n-4q} \leq c(\ell)$. If
$\frac{2n-q}{2n-4q}
> c(\ell)$, implying that $q > \frac{2-2c(\ell)}{1-4c(\ell)}n$, we use the lower bound of Lemma
\ref{lem:boundmin} to obtain
\[
\frac{2n-q}{ LB^{nt}_{mid}(n,\ell) } < \frac{2n-\frac{2-2c(\ell)}{1-4c(\ell)}n}{LB^{nt}_{mid}(n,\ell)} <
\frac{(2n-\frac{2-2c(\ell)}{1-4c(\ell)}n)\ell(\ell+3)}{2n(\ell +1)}= \frac{3\ell
c(\ell)}{4c(\ell)-1}\frac{\ell+3}{\ell+1}= c(\ell).
\]
The last equality follows directly since $(4c(\ell)-1)(\ell+1)=3\ell(\ell+3)$.\\
\end{proof}

\subsection{$PH$ and $MPPH$ where not all input genotypes are nontrivial}

Given an instance $G$ of $PH$ or $MPPH$ containing $n$ genotypes, $n_{nt}$ denotes the number of nontrivial
genotypes in $G$ and $n_t$ the
number of trivial genotypes; clearly $n = n_{nt} + n_t.$
\medskip
\begin{lemma}
\label{lem:cliquewith}
Let $G$ be an $n \times m$ instance of $PH(*,\ell)$, for some $\ell \geq 2$, where the compatibility
graph of the nontrivial genotypes in $G$ is a clique, $G$ is not equal to a single trivial genotype,
and no nontrivial genotype in $G$ is the sum of two trivial genotypes in $G$. Then at least
\[
\lbwith{n}{\ell} = \bigg \lceil \frac{n}{\ell} + 1 \bigg \rceil
\]
haplotypes are needed to resolve $G$.
\end{lemma}
\begin{proof}
Note that the lemma holds if $n_t \geq n/\ell + 1$. So we assume from now on that $n_t < n/\ell + 1$.

We first prove that the bound holds for $n_{nt} \leq \ell$. Combining this with $n_t < n/2 + 1$ gives that $n < 2\ell
+ 2$. Thus $n/\ell + 1 < 4$. Hence if $n_t \geq 4$ then we are done. Thus we only have to consider cases where both
$n_t \in \{0,1,2,3\}$ and $\ell \geq \max \{2,n_{nt}\}$. We verify these cases in Table \ref{tab:case}; note the
importance of the fact that no nontrivial genotype is the sum of two trivial haplotypes in verifying that these are
correct lower bounds. (Also, there is no $n_t = 1, n_{nt} = 0$ case because of the lemma's precondition.)
\begin{table}
\centering \caption{Case $n_t < 4$, $n_{nt}\leq \ell$ in proof of Lemma \ref{lem:cliquewith}} \label{tab:case}
\begin{tabular}{|c|c|c|}
\hline
$n_t$&$n_{nt}$&$\lceil n/\ell +1 \rceil$\\
\hline
0 & 1 & 2 \\
0 & $z \geq 2$ & $\leq \lceil z/z + 1 \rceil = 2$\\
1 & 1 & 2\\
1 & $z \geq 2$ & $\leq \lceil (z+1)/z + 1 \rceil = 3$\\
2 & 0 & 2\\
2 & 1 & $\leq 3$\\
2 & $z\geq 2$ & $\leq \lceil (z+2)/z + 1 \rceil = 3$\\
3 & 0 & $\leq 3$\\
3 & 1 & $\leq 3$\\
3 & 2 & $\leq 4$\\
3 & $z \geq 3$ & $\leq \lceil (z+3)/z + 1 \rceil = 3$\\
\hline
\end{tabular}
\end{table}

We now prove the lemma for $n_{nt} > \ell$. Note that in this case there exists a unique trivial haplotype $h_t$
consistent with all nontrivial genotypes. Suppose, by way of contradiction, that $N = N_t + N_{nt}$ is the size of the
smallest instance $G'$ for which the bound does not hold. Let $H$ be an optimal solution for $G'$ and let $h = |H|$.

Observe firstly that $N = 1$ (mod $\ell)$, because if this is not true we have that $\lbwith{N-1}{\ell} =
\lbwith{N}{\ell}$ and we can find a smaller instance for which the bound does not hold, simply by removing an
arbitrary genotype from $G'$, contradicting the minimal choice of $N$.

Similarly we argue that $h = \lbwith{N}{\ell}-1$, since if $h \leq \lbwith{N}{\ell}-2$ we could remove an arbitrary
genotype to yield a size $N-1$ instance and still have that $h < \lbwith{N-1}{\ell}$.

We choose a specific resolution of $G'$ using $H$ and represent it as a \emph{haplotype graph}. The vertices of this
graph are the haplotypes in $H$. For each nontrivial genotype $g \in G'$ there is an edge between the two haplotypes
that resolve it. For each trivial genotype $g \in G'$ there is a loop on the corresponding haplotype. There are no
edges between looped haplotypes because of the precondition that no nontrivial genotype is the sum of two trivial
genotypes.

From Lemma 5 of \cite{islands} it follows that, with the exception of the possibly present trivial haplotype and
disregarding loops, each haplotype in the graph has degree at most $\ell$. In addition, if an unlooped haplotype has
degree less than or equal to $\ell$, or a looped haplotype has degree (excluding its loop) strictly smaller than
$\ell$, then deleting this haplotype and all its at most $\ell$ incident genotypes creates an instance $G''$
containing at least $N-\ell$ genotypes that can be resolved using $h-1$ haplotypes, yielding a contradiction to the
minimality of $N$. (Note that, because $N_{nt}>\ell$, it is not possible that the instance $G''$ is empty or equal to
a single trivial genotype.)

The only case that remains is when, apart from the possibly present trivial haplotype, every haplotype in the
haplotype graph is looped and has degree $\ell$ (excluding its loop). However, there are no edges between looped
vertices and they can therefore only be adjacent to the trivial haplotype, yielding a contradiction.\\
\end{proof}
\medskip
\begin{lemma}
\label{lem:withgeneral}
Let $G$ be an $n \times m$ instance of $PH(*,\ell)$, for some $\ell \geq 2$, where $G$ is not equal to a
single trivial genotype, and no nontrivial genotype in $G$ is the sum of two trivial genotypes in $G$. Then
at least $\lbwith{n}{\ell}$ haplotypes are needed to resolve $G$.
\end{lemma}
\begin{proof}
Essentially the same inductive argument as used in Lemma \ref{lem:boundmin} works: it is always possible to disconnect
the compatibility graph of $G$ into at least two components by removing at most $\ell$ nontrivial genotypes, and using
cliques as the base of the induction. The presence of trivial genotypes in the input (which we can actually simply
exclude from the compatibility graph) does not alter the analysis. The fact that (in the inductive step) at least two
components are created, each of which contains at least one nontrivial genotype, ensures that the inductive argument
is not harmed by the presence of single trivial genotypes (for which the bound does not hold).\\
\end{proof}
\medskip
\begin{corollary}
\label{cor:withfirstbound} Let $G$ be an $n \times m$ instance of $PH(*,\ell)$ or $MPPH(*,\ell)$, for some $\ell
\geq 2$. Any feasible solution for $G$ is within a ratio of $2\ell$ from optimal.
\end{corollary}
\begin{proof}
Immediate because $2n/(n/\ell+1) < 2\ell$. (As before the algorithm from e.g. \cite{gusfieldlinear} can be used to
generate feasible solutions for $MPPH$, or to determine that they do not exist.)\\
\end{proof}
The algorithm $PH^{nt}M$ can easily be adapted to solve $PH(*,\ell)$ approximately.

\medskip

\noindent
\textbf{Algorithm:} $PHM$\\
\textbf{Step 1:} remove from $G$ all genotypes that are the sum of two trivial genotypes \\
\textbf{Step 2:} construct the compatibility graph $C(G')$ of the leftover instance $G'$.\\
\textbf{Step 3:} find a maximal matching $M$ in $C(G')$.\\
\textbf{Step 4:} for every edge $\{g_1,g_2\}\in M$, resolve $g_1$ and $g_2$ by
three haplotypes if $g_1$ and $g_2$ are both nontrivial and by two haplotypes if one of them is trivial.\\
\textbf{Step 5:} resolve each remaining nontrivial genotype by two haplotypes and each remaining trivial genotype by its corresponding haplotype.
\medskip
\begin{theorem}
$PHM$ computes a solution to $PH(*,\ell)$ in polynomial time within an approximation
ratio of $d(\ell)=\frac{3}{2}\ell +\frac{1}{2}$, for every $\ell \geq 2$.
\end{theorem}
\begin{proof}
Since constructing $C(G)$ given $G$ takes $O(n^2m)$ time and finding a maximal matching in any graph takes linear
time, $O(n^2m)$ running time follows directly.

Let $q$ be the size of the maximal matching, $n$ the number of genotypes
after Step 1 and $n_t$ the number of trivial genotypes in $G'$.
Then $PHM$
gives a solution with $2n-q-n_t$ haplotypes.
Since the complement of the
maximal matching is an independent set of size $n-2q$ in $C(G')$, any solution must contain at least $2(n-2q)$
haplotypes to resolve the genotypes in this independent set.
The theorem thus holds if $\frac{2n-q-n_t}{n-2q} \leq d(\ell)$.
If $\frac{2n-q-n_t}{n-2q}
> d(\ell)$, implying that $q > \frac{(d(\ell)-2)n+n_t}{2d(\ell)-1}$, we use the lower bound of Lemma
\ref{lem:withgeneral} and obtain
\[
\frac{2n-q-n_t}{LB_{mid}(n,\ell)} < \frac{2n-\frac{(d(\ell)-2)n+n_t}{2d(\ell)-1}}{\lceil \frac{n}{\ell} + 1 \rceil} <
\frac{2n-\frac{(d(\ell)-2)n}{2d(\ell)-1}}{\frac{n}{\ell}}= \frac{3d(\ell)\ell}{2d(\ell)-1}= d(\ell).
\]
The last equality follows directly since $2d(\ell)-1 = 3\ell$.\\
\end{proof}

\section{Postlude}
\label{sec:concl}
There remain a number of open problems to be solved. The complexity of $PH(*,2)$ and $MPPH(*,2)$ is still unknown. An
approach that might raise the necessary insight is to study the $PH(*,2)\text{-}Cq$ and $MPPH(*,2)\text{-}Cq$ variants
of these problems (i.e. where the compatibility graph is the sum of $q$ cliques) for small $q$. If a complexity result
nevertheless continues to be elusive then it would be interesting to try and improve approximation ratios for
$PH(*,2)$ and $MPPH(*,2)$; might it even be possible to find a PTAS (\emph{Polynomial-time Approximation Scheme}) for
each of these problems? Note also that the complexity of $PH(k,2)$ and $MPPH(k,2)$ remains open for constant $k \geq
3$.

Another intriguing open question concerns the relative complexity of $PH$ and $MPPH$ instances. Has $PH(k,\ell)$ always
the same complexity as $MPPH(k,\ell)$, in terms of well-known complexity measurements (polynomial-time solvability,
NP-hardness, APX-hardness)? For hard instances, do approximability ratios differ? A related question is whether it is possible to directly
encode $PH$ instances as $MPPH$ instances, and/or vice-versa, and if so whether/how this affects the bounds on the number
of 2's in columns and rows.

For hard $PH(k,\ell)$ instances it would also be interesting to see if those approximation algorithms that yield
approximation ratios as functions of $k$, can be intelligently combined with the approximation algorithms in this
paper (having approximation ratios determined by $\ell$), perhaps with superior approximation ratios as a consequence.
In terms of approximation algorithms for $MPPH$ there is a lot of work to be done because the
approximation algorithms presented in this paper actually do little more than return an arbitrary feasible solution.
It is also not clear if the $2^{k-1}$-approximation algorithms for $PH(k,*)$ can be attained (or improved) for $MPPH$.
More generally, it seems likely that big improvements in approximation ratios (for both $PH$ and $MPPH$) will require
more sophisticated, input-sensitive lower bounds and algorithms. What are the limits of approximability for these
problems, and how far will algorithms with formal performance-guarantees (such as in this paper) have to improve to
make them competitive with dominant ILP-based methods?

Finally, with respect to $MPPH$, it could be good to
explore how parsimonious the solutions are that are produced by the
various $PPH$ feasibility algorithms, and whether searching through
the entire space of $PPH$ solutions (as proposed in \cite{anOptimal})
yields practical algorithms for solving $MPPH$.

\section*{Acknowledgements}

All authors contributed equally to this paper and were supported by the Dutch BSIK/BRICKS project. A preliminary
version of this paper appeared in \emph{Proceedings of the 6th International Workshop on Algorithms in Bioinformatics} (WABI 2006)
\cite{wabibeaches}.

\clearpage

\begin{biography}{Leo van Iersel}
received in 2004 his Master of Science degree in Applied Mathematics from the Universiteit Twente in The Netherlands.
He is now working as a PhD student at the Technische Universiteit Eindhoven, also in the Netherlands. His research is
mainly concerned with the search for combinatorial algorithms for biological problems.
\end{biography}

\begin{biography}{Judith Keijsper}
received her master's and PhD degrees in 1994 and 1998 respectively from the Universiteit van Amsterdam in The
Netherlands, where she worked with Lex Schrijver on combinatorial algorithms for graph problems. After working as a
postdoc at Leibniz-IMAG in Grenoble, France, and as an assistant professor at the Universiteit Twente in the
Netherlands for short periods of time, she moved to the Technische Universiteit Eindhoven in the Netherlands in the
year 2000. She is an assistant professor there, and her current research focus is combinatorial algorithms for
problems from computational biology.
\end{biography}

\begin{biography}{Steven Kelk}
received his PhD in Computer Science in 2004 from the University
of Warwick, in England. He is now working as a postdoc at the
Centrum voor Wiskunde en Informatica (CWI) in Amsterdam, the
Netherlands, where he is focussing on the combinatorial aspects of
computational biology.
\end{biography}

\begin{biography}{Leen Stougie}
received his PhD in 1985 from the Erasmus Universiteit of Rotterdam, The Netherlands. He is currently working at the
Centrum voor Wiskunde en Informatica (CWI) in Amsterdam and at the Technische Universiteit Eindhoven as an associate
professor.
\end{biography}

\end{document}